\documentclass[twocolumn,prl,superscriptaddress]{revtex4}
\usepackage{amsmath,amssymb,amsfonts,mathrsfs}
\usepackage{amsthm}
\usepackage{graphicx}
\usepackage{dcolumn}
\usepackage{bm}
\usepackage{color}

\begin{document}
	
	\title{Note on \\ ``Experimental Measurement of Quantum Metric Tensor and Related Topological Phase Transition with a Superconducting Qubit"}
\author{Yan-Qing Zhu}
\affiliation{National Laboratory of Solid State Microstructures, School of Physics,
Nanjing University, Nanjing 210093, China}

\author{Dan-Wei Zhang}
\affiliation{Guangdong Provincial Key Laboratory of Quantum Engineering and Quantum Materials, GPETR Center for Quantum Precision Measurement and SPTE, South China Normal University, Guangzhou 510006, China}

\author{Xinsheng Tan}
\affiliation{National Laboratory of Solid State Microstructures, School of Physics,
Nanjing University, Nanjing 210093, China}

\author{Hai-Feng Yu}
\affiliation{National Laboratory of Solid State Microstructures, School of Physics,
Nanjing University, Nanjing 210093, China}

\author{Hui Yan}
\affiliation{Guangdong Provincial Key Laboratory of Quantum Engineering and Quantum Materials, GPETR Center for Quantum Precision Measurement and SPTE, South China Normal University, Guangzhou 510006, China}

\author{Yang Yu}
\affiliation{National Laboratory of Solid State Microstructures, School of Physics,
Nanjing University, Nanjing 210093, China}

\author{Shi-Liang Zhu}
\affiliation{National Laboratory of Solid State Microstructures, School of Physics,
Nanjing University, Nanjing 210093, China}
\affiliation{Guangdong Provincial Key Laboratory of Quantum Engineering and Quantum Materials, GPETR Center for Quantum Precision Measurement and SPTE, South China Normal University, Guangzhou 510006, China}

\begin{abstract}
In the paper [X. Tan \textit{et al.}, Phys. Rev. Lett.
122, 210401 (2019)], we have studied the Euler characteristic number $\chi$ for a two-band model; However, the $\chi$ calculated there is not an integer when the parameter $|h|>1$ and then may not be considered as a suitable topological number for the system. In this note, we find that the Bloch vectors of the ground state for $|h|>1$ do not cover the whole Bloch sphere and thus the Euler characteristic number should be effectively associated with the manifold of a disk, rather than a sphere. After taking into account the boundary contribution, we derive the correct Euler characteristic number $\chi$. Unfortunately, the Euler characteristic number $\chi$ does not change when crossing the critical points $|h|=1$ and thus can not be used to characterize the topological phase transition of the present model.
\end{abstract}
%``Euler characteristic number" $\chi$

\maketitle

%	\section{Quantum geometric tensor}
	
One of the Hamiltonians studied in our paper \cite{XTan2019} and in Ref. \cite{YQMa2013} is given by
	\begin{equation}
	H_2(k_x,k_y)=\frac{\Omega}{2}
	\begin{pmatrix}
	-h-\cos k_x   &   \alpha\sin k_x e^{-ik_y}\\
	\alpha\sin k_x e^{ik_y}   & h+\cos k_x
	\end{pmatrix}, \label{Ham2}
	\end{equation}
where the parameters $\boldsymbol k=(k_x,k_y)$ with $\{k_x,k_y\}\in[0,2\pi]$ in Ref. \cite{XTan2019} and with $k_x \in[0,2\pi]$ and $k_y \in[0,\pi]$ in Ref. \cite{YQMa2013}. The Hamiltonian (\ref{Ham2}) can be rewritten as
 \begin{equation}
 H_2(\boldsymbol k)=d_x\sigma_x+d_y\sigma_y+d_z\sigma_z,
 \end{equation}
where the Bloch vectors $\mathbf{d}=(d_x,d_y,d_z)$ are given by
\begin{eqnarray}
d_x &=&(\alpha\Omega/2)\sin k_x\cos k_y,\\
d_y&=& (\alpha\Omega/2)\sin k_x\sin k_y,\\
d_z&=&-\Omega(h+\cos k_x)/2.
   \end{eqnarray}
   The energy spectrum are obtained as $E_{\pm}=\pm d$ with $d=\sqrt{d_x^2+d_y^2+d_z^2}$ and the corresponding eigenvectors are given by
  % \begin{equation} |u_\pm\rangle= \begin{pmatrix} d_x-id_y \\ \pm d-d_z\end{pmatrix}\end{equation}

  \begin{equation}
	|u_\pm\rangle=\frac{\Omega}{2}
	\begin{pmatrix}
	d_x-id_y \\
	\pm d-d_z
	\end{pmatrix}.
	\end{equation}

We first introduce the method addressed in Ref. \cite{YQMa2013} to calculate  the Euler characteristic number $\chi$ of a two-level (two-band) model with the two-dimensional coordinates  $({\mu,\nu})$, it is given by
 \begin{equation}
\chi=\frac{1}{4\pi}\int_{\mathcal{M}} R \sqrt{\det g}~d\mu d\nu, \label{EulerNum}
\end{equation}
where $R$ is the Ricci scalar curvature and $g$ denotes the $2\times2$ metric tensor. In a two-level gapped system, it is generally true that $\sqrt{\det{g}}=|\mathcal{F}_{\mu\nu}|/2$, where the Berry curvature $\mathcal{F}_{\mu\nu}=\hat{\mathbf{d}}\cdot(\partial_{\mu}\hat{\mathbf{d}}\times\partial_{\nu}\hat{\mathbf{d}})/2$ with $\hat{\mathbf{d}}=\mathbf{d}/d$.
Associated with the Berry curvature, we have the first Chern number defined as
\begin{equation} C=\frac{1}{2\pi}\int\mathcal{F}_{\mu\nu} d\mu d\nu. \label{Chern}\end{equation}

As for the Hamiltonian in Eq. (\ref{Ham2}), the quantum metric for both the ground and excited states is obtained as
	\begin{equation}
	g=\frac{1}{4}
	\begin{pmatrix}
	\alpha^2(1+h\cos k_x)^2/f^2 & 0 \\
	0 & \alpha^2\sin^2 k_x/f
	\end{pmatrix},\label{Gm}
	\end{equation}
where $f=(h+\cos k_x)^2+\alpha^2\sin^2 k_x.$
One can find that the quantum metric tensor $g$ is independent on $k_y$ and the square root of its determination
\begin{equation}\label{detg}
\sqrt{\det g}=\alpha^2|(1+h\cos k_x)\sin k_x|/(4f^{3/2}).
\end{equation}
Please note that there is a typo in Eq. (7) in Ref. \cite{XTan2019}, that is, the denominator $2f^{3/2}$ there should be $4f^{3/2}$ as corrected in Eq. (\ref{detg}). The Berry curvature of the ground state is given by
\begin{equation}
\mathcal{F}=\alpha^2(1+h\cos k_x)\sin k_x/(2f^{3/2}) \label{Bcurvature}.
\end{equation}

In Ref. \cite{XTan2019}, we parameterize the system with the spherical coordinate ($\theta$,$\phi$) with the notations $k_x \rightarrow \theta$, $k_y \rightarrow \phi$, and the Euler characteristic number $\chi$ is calculated by substituting Eq. (\ref{Gm}) and the Ricci scalar curvature $R=8$ (the Gaussian curvature $K=4$)
of the quantum state manifold into Eq. (\ref{EulerNum}). While the Chern number $C$ can be obtained by substituting Eq. (\ref{Bcurvature}) into Eq. (\ref{Chern}). The results are plotted in Fig. 1 (Fig. 4(b) in Ref. \cite{XTan2019}). It shows that the $C$ is always zero; while the $\chi=4$ when $|h|<1$, but it just gradually becomes zero when $|h|$ approaches infinite. Similar results of $\chi$ are also demonstrated in Ref. \cite{YQMa2013}.

\begin{figure}[http]\centering
\includegraphics[width=0.9\columnwidth]{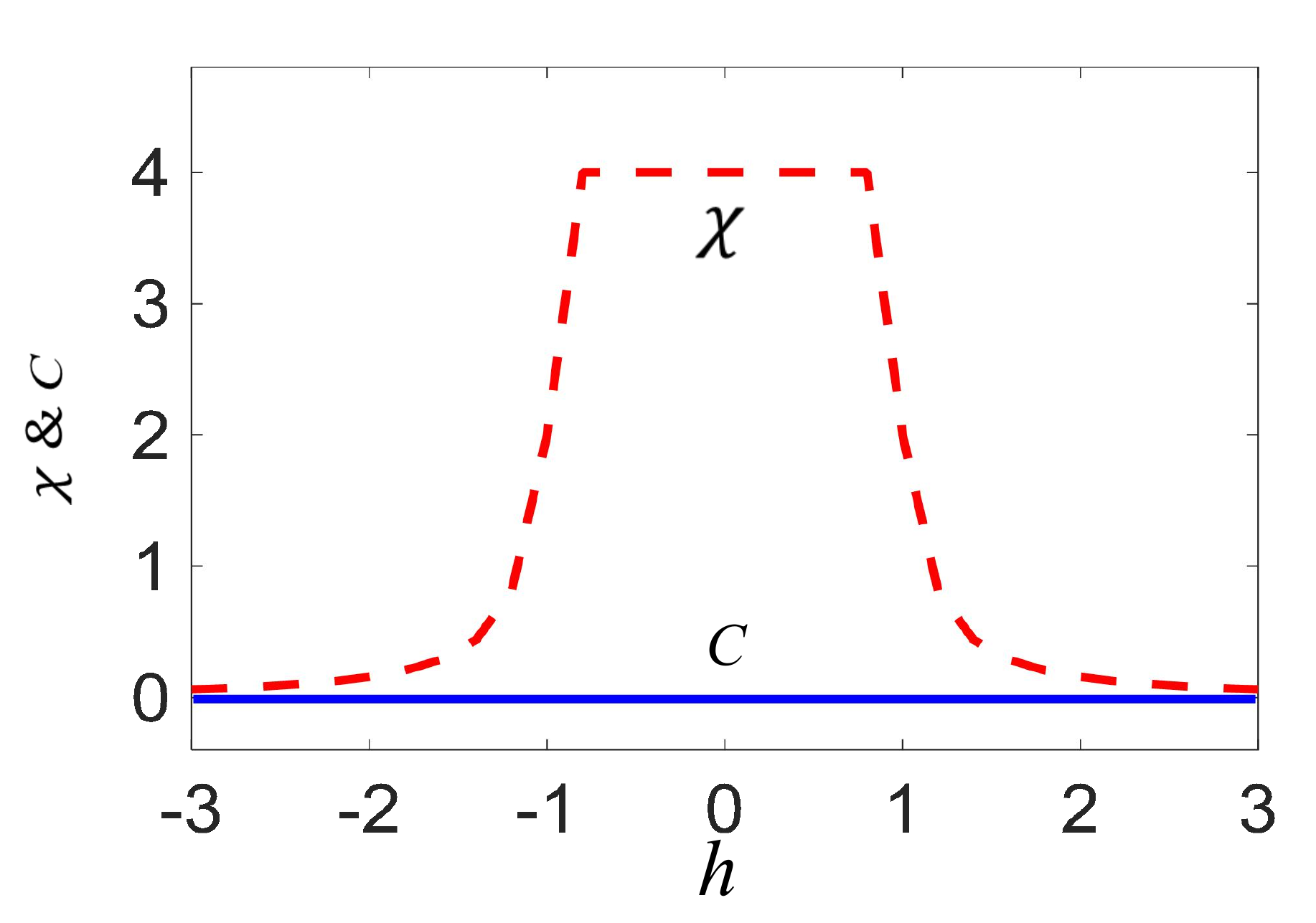}
\caption{The Chern number $C$ and ``Euler characteristic number" $\chi$  calculated in Refs. \cite{XTan2019,YQMa2013} as a function of the parameter $h$.  }
\label{Top}
\end{figure}

We here present a physics picture to illustrate the results plotted in Fig. 1.
For simplicity, we take $\alpha=1$ (and $\Omega/2$ as the energy unit) as an example to illustrate the physical meanings of the results. Under these conditions, the Bloch vectors $\mathbf{d}$ for a fixed $h$ form a unit sphere with the center of it at $(0,0,h)$, and for concreteness, we denote the sphere formed by $\mathbf{d}$ as $S_d$. We may assume that there is a monopole at the origin $\mathbf{d}=(0,0,0)$, as shown in Fig. 2. We can see that the sphere $S_d$ contains the coordinate origin $\mathbf{d}=0$ when $|h|<1$, while it does not contain the origin point when $|h|>1$.

The Chern number $C$ is determined by the sum of the net Berry flux flowed past the surface of the sphere $S_d$; In contrast, the ``Euler characteristic number" $\chi$ defined in Eq.(\ref{EulerNum}) is determined by the sum of the \textit{absolute} Berry flux flowed past the surface of the sphere $S_d$.

As for the Chern number $C$, it should be 1 for the case in Fig. 2(a) and 0 in Fig. 2(b). However, as for the Hamiltonian in Eq. (1), the Bloch vectors $\mathbf{d}$ run over the sphere $S_d$ twice (the Bloch sphere is defined as a unit sphere with the center of it at the origin, but in general, it is not the sphere $S_d$ formed by the Bloch vectors $\mathbf{d}$), and the topological charge of the monopole at the origin is opposite [that is, there exists a pair of monopole and anti-monopole at the origin for $k_x \in (\pi,2\pi)$ and $k_y \in (0,2\pi)$)]. So the Chern number $C$ of the system is always zero for all $h$, which is consistent with the result in Fig. 1.

\begin{figure}[http]\centering
\includegraphics[width=0.9\columnwidth]{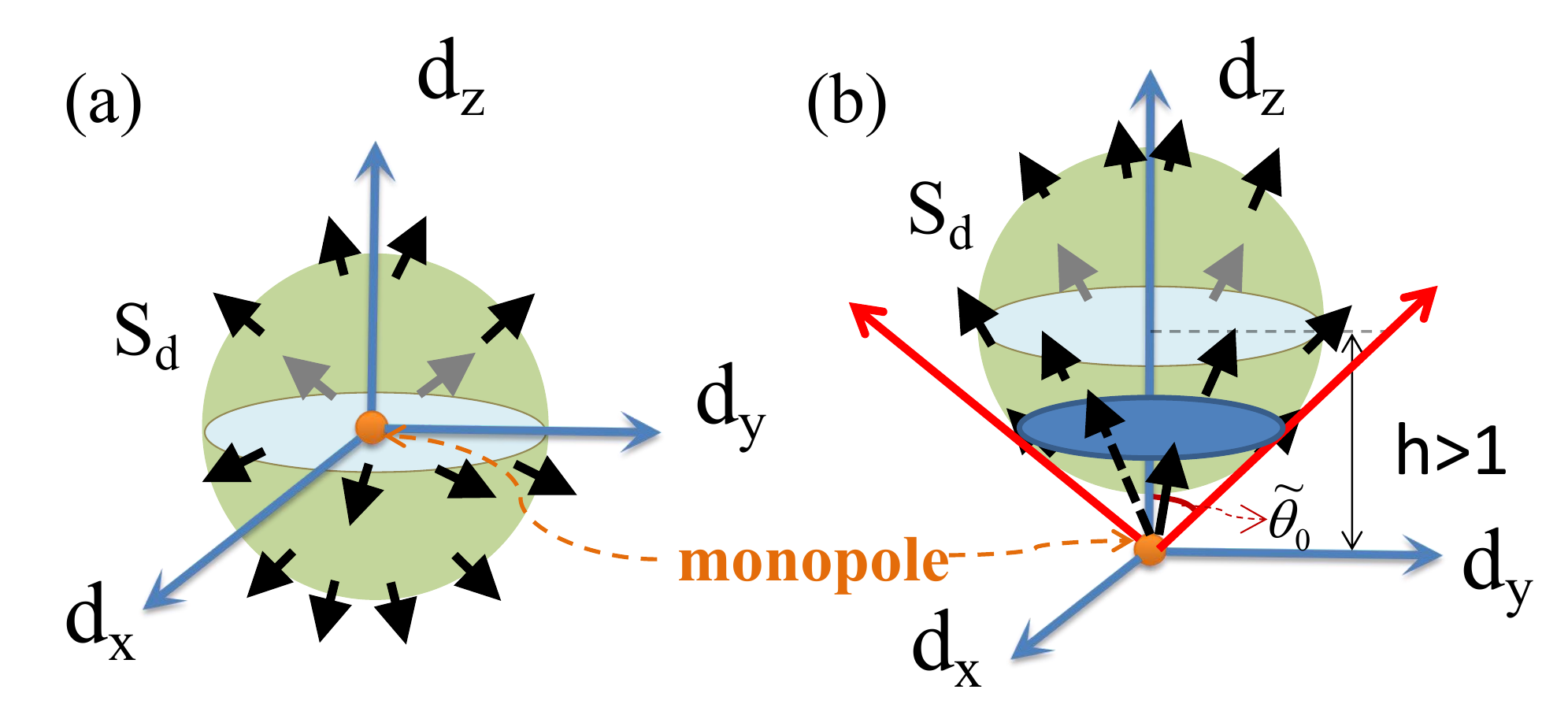}
\caption{ Schematic diagram of the Bloch vectors $\mathbf{d}$ in the region (a) $|h|<1$ (here $h=0$) and (b) $|h|>1$, respectively. It shows the
topological difference between the two distinct phases when
the sphere $S_d$ moves from the degeneracy in the $z$
direction by distance $h$. The Berry flux vectors are schematically
presented by arrows. The Chern number $C$ is determined by the sum of the net flux flowed past the surface of the sphere $S_d$; In contrast,  the ``Euler characteristic number" $\chi$ calculated in Refs. \cite{XTan2019,YQMa2013} is determined by the sum of the \textit{absolute} flux flowed past the surface of the sphere $S_d$.  }
\label{BS}
\end{figure}

As for the ``Euler characteristic number" $\chi$ calculated in Refs. \cite{XTan2019,YQMa2013}, it is actually determined by the solid angle of the surface formed by the unit $\hat{\mathbf{d}}$. If we parameterize and normalize the Bloch vectors with the spherical coordinate ($\tilde{\theta}$, $\phi$) as
\begin{equation}
\begin{aligned}\label{BloVector}
\hat{d}_x&=d_x/d=\sin\tilde{\theta}\cos\phi,\\
\hat{d}_y&=d_y/d=\sin\tilde{\theta}\sin\phi,\\
\hat{d}_z&=d_z/d=\cos\tilde{\theta}.
\end{aligned}
\end{equation}
The ``Euler characteristic number" we obtain now is given by \cite{XTan2019,YQMa2013}
\begin{equation}
\begin{aligned}
\chi=\frac{1}{2\pi}\int_{\mathcal{M}} KdA = \frac{2}{\pi}\int \sqrt{\det g}d\tilde{\theta} d\phi\\
=\frac{1}{\pi}\int |\mathcal{F}_{\tilde{\theta}\phi}|d\tilde{\theta} d\phi=\frac{1}{2\pi} \int d\Omega,
\end{aligned}
\end{equation}
where $d\Omega=|\hat{\mathbf{d}}\cdot\partial_{\tilde{\theta}}\hat{\mathbf{d}}\times\partial_{\phi}\hat{\mathbf{d}}|d\tilde{\theta} d\phi=\sin\tilde{\theta} d\tilde{\theta} d\phi$. Figure \ref{BS} also shows the solid angle of these two different situations. Figure \ref{BS}(a) shows the situation for $|h|<1$, the solid angle $\Omega$ is $2\times4\pi$, thus we obtain $\chi=4$; when $|h|$ becomes larger and larger in the parameter region $|h|>1$, the sphere $S_d$ is getting farther away from the origin along the $z$ axis and thus the solid angle formed by $\hat{\mathbf{d}}$ is $2\times 4\pi(1-\cos\tilde{\theta}_0)$ and becomes smaller and smaller until zero, as illustrated in Fig. \ref{BS}(b). Therefore, $\chi$ continuously tends from 4 at $|h|=1$ to zero at the infinity $|h|\rightarrow\infty$.

However, the above results imply that the calculation of the ``Euler characteristic number"  for the two-level system must be not correct for $|h|>1$ since it is not an integer. In the following, we will show that when $|h|>1$, the above calculation misses the boundary contribution since the Bloch vectors of the ground state do not cover the whole Bloch sphere (that is, the unit vector $\hat{\mathbf{d}}$ only contains part of the Bloch sphere). In this case, the Euler characteristic number should be associated with the manifold of a disk, rather than a sphere. After taking into account the boundary contribution, we can derive the correct Euler characteristic number $\chi$.

\begin{figure}[http]\centering
\includegraphics[width=0.9\columnwidth]{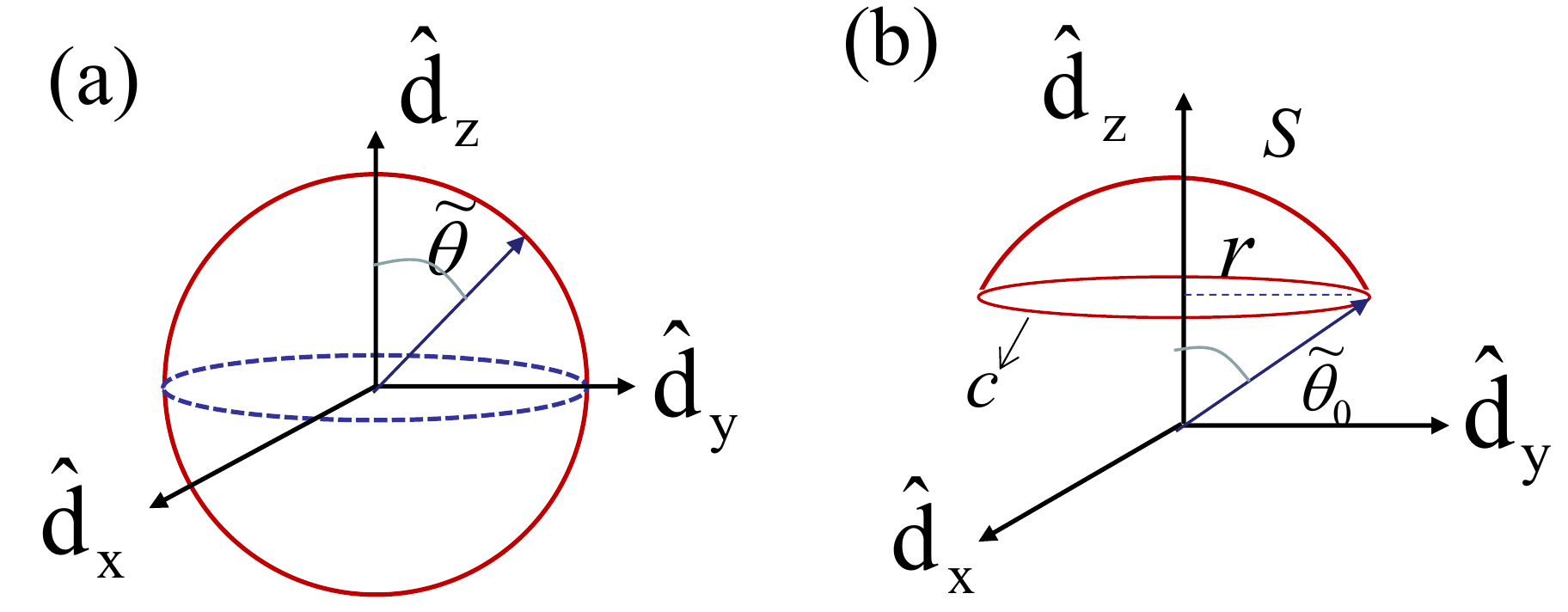}
\caption{ Schematic diagram of the surface formed by the unit $\hat{\mathbf{d}}$. (a) the unit $\hat{\mathbf{d}}$  formed a unit Bloch sphere for $|h|<1$. (b) For $|h|>1$, the unit $\hat{\mathbf{d}}$ just formed a part of unit Bloch sphere $S$ with boundary $c$ , where $r=\sin\tilde{\theta}_0$ is the radius of $c$.
In our model, the  $\hat{\mathbf{d}}$ runs over (a) the unit Bloch sphere twice and (b) a part of unit Bloch sphere $S$ four times, thus we always obtain $\chi=4$. }
\label{MF}
\end{figure}

In mathematics, the Fubini-Study metric is a K$\ddot{a}$hler metric on projective Hilbert space $P\mathcal{H}$, namely, on a complex projective space $CP^n$, which is a space of complex lines through the origin of a $(n+1)$-dimensional complex vector space. In quantum mechanical systems, the metric $g_{\mu\nu}$ defined above measures the distance of ``rays" in projective Hilbert space $P\mathcal{H}=\mathcal{H}\setminus\{0\}/U(1)$. Recall that a ray as a point on manifold $\{|\tilde{\psi}\rangle\}$ is an equivalence class $[|\psi\rangle]$ (denoted by $|\tilde{\psi}\rangle$) of states $|\psi\rangle$ ($|\psi\rangle\neq0$) in Hilbert space $\mathcal{H}$, which differ only in phase \cite{Provost,Grigorenko}. For the $(n+1)$-dimensional Hilbert space $\mathcal{H}\setminus \{0\}=S^{2n+1}$, the manifold $\{|\tilde{\psi}\rangle\}$ is the projective Hilbert space $P\mathcal{H}=S^{2n+1}/U(1)=CP^{n}$.  Therefore, the topological number for the manifold of quantum states actually is the Euler characteristic number of $CP^n$. For a two-dimensional Hilbert space, its projective Hilbert space $P\mathcal{H}=S^3/S^1\backsimeq S^2$ is a Bloch sphere. So its Euler characteristic number $\chi=2$ and its restriction to the real tangent bundle should yield an expression of the ordinary ``round metric" of radius $\rho=1/2$ (and $K=1/\rho^2=4$) on $S^2$.

In our two-level system, the Euler characteristic number should be 4 for $|h|<1$ since the unit Bloch vectors $\hat{\mathbf{d}}$ run over the Bloch sphere twice (since $\{k_x,k_y\}\in[0,2\pi]$ ) and it is independent of $h$.
%From these points of view, the Gaussian curvature $K=4$ calculated by the quantum states is reasonable and correct,
However, the ``Euler characteristic number" obtained above in the region $|h|>1$ is not an integer and thus is not a correct one.
The reason is the following: the effective integral surface consisting of the eigenvetors $\{|{u}_-\rangle\}$ (more precisely, the states $\{|\tilde{u}_-\rangle\}$ in the projective Hilbert space) does not cover the entire Bloch sphere when $|h|>1$. As shown in Fig. 3(a) when $|h|<1$, $\hat{\mathbf{d}}$ runs over the unit Bloch sphere twice, and thus $\chi=4$. When $|h|>1$, $\hat{\mathbf{d}}$ runs over just a part of the unit Bloch sphere four times [Fig. 3(b)]. Note that the manifold now has a boundary. Under this condition, Eq. (\ref{EulerNum}) for calculating the Euler characteristic number is not correct since it misses the contribution of the boundary.

As for a manifold with a boundary, the Gauss-Bonnet formula can be generalized as
\begin{equation}\label{GBF}
\chi=\chi_{\text{bulk}}+\chi_{\text{boundary}}=\frac{1}{2\pi}\left(\int_\mathcal{M} KdA+\int_{\partial \mathcal{M}}k_g d l\right),
\end{equation}
where $\chi_{\text{bulk}}$ and $\chi_{\text{boundary}}$ are the bulk and boundary contributions
to the Euler characteristic number of the manifold $\mathcal{M}$ \cite{Kolodrubetz}. Here $k_g$ is the geodesic curvature of the boundary $\partial M$, and $d l$ is the line element along the boundary of $\mathcal{M}$. The metric is written in its fundamental form as
\begin{equation}
ds^2 = g_{11}d\lambda_1 + 2g_{12} d\lambda_1d\lambda_2 + g_{22}d\lambda_{2}d\lambda_{2}.
\end{equation}
These invariants are given by \cite{Kreyszig}
\begin{equation}
\begin{aligned}
k_g&=-\Gamma^1_{22}\frac{\sqrt{\det g}}{g_{22}^{3/2}},\\
dl&=\sqrt{g_{22}}d\lambda_2,
\end{aligned}
\end{equation}
where $k_g$ and $dl$ are given for a curve of constant $\lambda_1$, The Christoffel symbols $\Gamma^k_{ij}$ are defined as
\begin{equation}
\Gamma^k_{ij}=\frac{1}{2}g^{km}(\partial_jg_{im}+\partial_ig_{jm}-\partial_mg_{ij}),
\end{equation}
where $g^{ij}$ is the inverse of the metric tensor $g_{ij}$. The metric tensor in the spherical coordinate with $(\tilde{\theta},\phi) \leftrightarrow (\lambda_1,\lambda_2)$ takes the form \cite{XTan2019},
\begin{equation}
\begin{aligned}
g=\begin{pmatrix} g_{\tilde{\theta}\tilde{\theta}}& g_{\tilde{\theta}\phi}\\
g_{\phi\tilde{\theta}} & g_{\phi\phi}
\end{pmatrix}=\begin{pmatrix} 1/4 & 0\\
0 & \sin^2\tilde{\theta}/4
\end{pmatrix}.
\end{aligned}
\end{equation}
Then one can obtain that $\sqrt{\det g}=\sin\tilde{\theta}/4$, $\Gamma^1_{22}=-\frac{1}{2}g^{11}\partial_{1}g_{22}=-\sin\tilde{\theta}\cos{\tilde{\theta}}$, and thus we have $k_g=2\cos\tilde{\theta}_0/\sin{\tilde{\theta}_0}$, $dl=\sin\tilde{\theta}_0d\phi/2$ with constant $\tilde{\theta}_0$ on the boundary $\partial \mathcal{M}$. Note that $k_g$ calculated from the quantum metric $g_{\tilde{\theta}\phi}$ is the exact geodesic curvature of a circle with radius $r=\sin\tilde{\theta}_0/2$ on a sphere of radius $\rho=1/2$, where $k_g=\sqrt{\rho^2-r^2}/\rho r$ in the geometrical representation. For convenience, we consider the base manifold $S$ formed by $\hat{\mathbf{d}}$ (i.e., $K=1$ and $\rho=1$) with a circle boundary $c$. Thus, the Euler characteristic number in the region $|h|>1$ is given by
\begin{equation}
\begin{aligned}
\chi&=4\times\frac{1}{2\pi}\big(\int_S\sin{\tilde{\theta}}d\tilde{\theta}d\phi+\int_c
 \cos\tilde{\theta}_0d\phi\big)\\
 &=4\times\big(\int_0^{\tilde{\theta}_0}\sin\tilde{\theta}d\tilde{\theta}+\cos\tilde{\theta}_0\big)=4.
 \end{aligned}
\end{equation}
This result can be intuitively understood as the manifold of quantum states $\{|\tilde{u}_-\rangle\}$ in this case is equivalent to four disks due to $\hat{\mathbf{d}}$ runs over the surface $S$ four times, as depicted in Fig. 3(b). The two-dimensional compact surface $S$ with the circle boundary $c$ is topologically equivalent to a disk and thus has the Euler characteristic number $\chi(S)=1$, one can easily check this result for $S$ by using Eq. (\ref{GBF}).

In summary, the Euler characteristic number for our two-level quantum system should be always $4$ after we consider the contribution of boundary for the manifold of quantum states (formed by $\hat{\mathbf{d}}$) in $CP^1$ (unit Bloch sphere) in the region $|h|>1$. Unfortunately, the Euler characteristic number $\chi$ does not change when crossing the critical points $|h|=1$ and thus can not be used to characterize the topological phase transition of the present model.

We will submit an Erratum to PRL based on this note.

%Therefore, the ``Euler number" calculated in Refs. \cite{XTan2019,YQMa2013} is not an integer when $|h|>1$ and then may not be considered as a suitable topological number for the system.
	
	\vspace{0.5cm}
{\sl Acknowledgements}: We thank Yu-Quan Ma, Giandomenico Palumbo, and Nathan Goldman for helpful discussions.


\begin{thebibliography}{99}
		
	
\bibitem{XTan2019} X. Tan, D.-W. Zhang, Z. Yang, J. Chu, Y.-Q. Zhu,
D. Li, X. Yang, S. Song, Z. Han, Z. Li, Y. Dong, H.-F.
Yu, H. Yan, S.-L. Zhu, and Y. Yu, Experimental Measurement of Quantum Metric Tensor and Related Topological Phase Transition with a Superconducting Qubit, Phys. Rev. Lett.
122, 210401 (2019).
	
%\bibitem{Ma2010} Y.-Q. Ma, S. Chen, H. Fan, and W.-M. Liu, Abelian and non-Abelian quantum geometric tensor, \newblock {\it Phys. Rev. B} {\bf81}, %245129 (2010).
		

\bibitem{YQMa2013} Y.-Q. Ma, S.-J. Gu, S. Chen, H. Fan, and W.-M. Liu, The Euler number of Bloch states manifold and the quantum
phases in gapped fermionic systems, Europhys. Lett. {\bf103}, 10008 (2013).
		
\bibitem{Provost} J. P. Provost and G. Vallee, Riemannian structure on manifolds
of quantum states, Commun. Math. Phys. {\bf76}, 289 (1980).

\bibitem{Grigorenko} A. N. Grigorenko, Geometry of projective Hilbert space, Phys. Rev. A {\bf46}, 7292 (1992).

\bibitem{Kolodrubetz} M. Kolodrubetz, V. Gritsev, and A. Polkovnikov, Classifying and measuring geometry of a quantum ground state
manifold, Phys. Rev. B {\bf88}, 064304 (2013).

\bibitem{Kreyszig} E. Kreyszig, \emph{Differential Geometry} (University of Toronto Press,
Toronto, 1959).
		
\end{thebibliography}
\end{document}